\documentclass[aps,prd,epsf,showpacs,amsmath,amssymb,graphics,
twocolumn,10pt]{revtex4}
\usepackage{graphicx}
\usepackage{dcolumn}
\usepackage{bm}

\newcommand{\be}{\begin{equation}}
\newcommand{\ee}{\end{equation}}
\newcommand{\ba}{\begin{eqnarray}}
\newcommand{\ea}{\end{eqnarray}}
\newcommand{\ban}{\begin{eqnarray*}}
\newcommand{\ean}{\end{eqnarray*}}

\newcommand{\eq}[1]{(\ref{#1})}

\begin{document}

\title{Ultra-high energy particle collisions in a regular spacetime
without blackholes or naked singularities}

\author{Mandar Patil \footnote{ Electronic address: mandarp@tifr.res.in}
and  Pankaj S. Joshi \footnote{ Electronic address: psj@tifr.res.in}}

\affiliation{Tata Institute of Fundamental Research\\
Homi Bhabha Road, Mumbai 400005, India}

\begin{abstract}
We investigate here the particle acceleration and collisions
with extremely large center of mass energies in a perfectly regular
spacetime containing neither singularity nor an event horizon.
The ultra-high energy collisions of particles near the event horizon
of extremal Kerr blackhole, and also in many other examples of
extremal blackholes have been investigated and reported recently.
We studied an analogous particle acceleration process in the
Kerr and Reissner-Nordstrom spacetimes without horizon, containing
naked singularities. Further to this, we show here that the
particle acceleration and collision process is in fact independent
of blackholes and naked singularities,
and can happen in a fully regular spacetime containing neither
of these. We derive the conditions on the general static spherically
symmetric metric for such a phenomena to happen. We show that
in order to have ultra-high energy collisions it is necessary
for the norm of the timelike Killing vector to admit a maximum with
a vanishingly small but a negative value. This is also a condition
implying the presence of a surface with extremely large but
nevertheless finite value of the redshift or blueshift.
Conditions to have ultrahigh energy collisions and regular center
imply the violation of strong energy condition near the center
while the weak energy condition is respected in the region around 
the center. Thus the central region is surrounded by a dark energy 
fluid. Both the energy conditions are respected at the location 
where the high energy collisions take place.
As a concrete example we then investigate the acceleration
process in the spacetime geometry derived by Bardeen which is
sourced by a non-liner self-gravitating magnetic monopole.
\end{abstract}
\maketitle

\section{Introduction}
Various particle accelerators like the Large Hadron Collider
can accelerate and collide the particles upto 10Tev.
Physics beyond this scale, all the way upto Planck scale which
is at $10^{19}$GeV remains unexplored experimentally.
An intriguing possibility is to explore the naturally occuring
astrophysical phenomena where such collisions can take
place, and to extract information about the new physics from
the signals that we get from such collision events.

In this direction it was suggested by Banados, Silk and West
\cite{BSW}
that extremal or near extremal
Kerr blackholes can serve the purpose. They showed that it would
be possible to have collisions with extremely large center
of mass energies in the vicinity of the event horizon of the
extremal Kerr blackholes, even if the colliding particles
start out from rest at infinity. For such collisions to occur
it is necessary, however,  to have the geodesic parameters of
one of the colliding particles highly fine-tuned and the proper
time required for the collisions in the rest frame
of this particle also turns out to be infinite. Various other
drawbacks of this process, when it is explored in the context
of realistic astrophysical scenario were discussed in
\cite{Berti}.
Different aspects of particle acceleration mechanism
for Kerr blackhole were investigated subsequently
\cite{Harada}.
The emergent flux of the particles like neutrinos created
in the collisions of dark matter particles in the Kerr blackhole
geometry around the event horizon was computed and shown to
be observable with detectors like Icecube
\cite{BSW2}.
It was also argued that the maximum center of mass energy
of collision achievable, taking into account the backreaction,
might be actually significantly smaller than the Planck scale
\cite{Nakao}.
This process was also shown to occur in various other
extremal blackhole geometries
\cite{Others}.

We showed that the process of ultrahigh energy collisions
goes beyond the blackhole geometries and can also be extended
to the spacetimes having no event horizon but containing
the naked singularities.
We explored the Kerr and Reissner-Nordstrom geometries from
this perspective. We showed that it is possible to have high energy
collisions in the Kerr geometry with spin parameter exceeding
unity by a vanishingly small amount. The event horizon is absent
in this case and the singularity is exposed to the asymptotic
observers. The high energy collisions take place at a location
faraway from the singularity, where the horizon would have
been present in the extremal Kerr blalckhole case. We considered
the particles that follow the geodesic motion in the equatorial
plane as well as along the axis of symmetry
\cite{Patil}.
Due to the absence of the event horizon collisions can
take place between the radially ingoing and outgoing particles
unlike in the blackhole case where one is forced to consider
the collisions between two ingoing particles only. We showed that
this allows us to avoid the extreme finetuning of the geodesic
parameters and that the proper time required for the collisions
to take place also happens to be finite. Interestingly, the
repulsive effect of gravity near the naked singularity plays
an important role near the axis of symmetry in the particle
collision process. We also explored the Reissner-Notdstrom
geometry from the perspective of the particle acceleration and
performed an exact calculation computing the center of mass
energy of collision taking into account the backreaction effect
of the colliding particles on the background geometry. This
can be done by considering the collision between the spherical
shells of particles instead of individual particles themselves
\cite{Patil2}.
As we showed, the maximum center of mass energy of collision
achievable can be many orders of magnitude larger than the Planck
energy. We also showed that, on an average, about half of the
particles which are produced in the high energy collisions would
escape to infinity, essentially due to the absence of the event
horizon and thus would give rise to a large flux of emergent particles.
\cite{Patil3}.
On the other hand, in the blackhole case, due to the presence
of the event horizon, the emergent flux would be small as most
of the particles generated in the collisions are captured
by the event horizon.

We also showed that the ultrahigh energy collisions between the
ingoing particles can take place in the vicinity of the Janis-Newmann-Winicor
naked singularities, obtained by generalizing Schwarzschild metric by
invoking a massless scalar field, while such collisions were absent
in the Schwarzschild blackhole geometry\cite{Patil4}.

In the present work, we go one step ahead and show that
it would be possible to have ultrahigh collisions in the absence
of both blackholes as well as naked singularities. This is
quite surprising because based on the intuition it might be
thought that it would be necessary to have either blackholes or
singularities in order to accelerate the particles and
make them collide with large center of mass energies.

For the sake of simplicity and definitiveness we focus
here on the spherically symmetric static spacetimes. We derive
the conditions necessary to have high energy particle collisions
and at the same time for the spacetime to be perfectly regular
and devoid of any singularity and event horizon.
As an concrete example we discuss then the spacetime metric
proposed by Bardeen
\cite{Bardeen}.
It is a two-parameter solution, the parameters being the mass
and the charge of the geometry. When the mass parameter is larger
than a certain critical value, the spacetime contains a blackhole
which has a regular center and which is interestingly free from
a central singularity. When the mass parameter is smaller
than a critical value the spacetime does not contain an event
horizon and is still regular everywhere. It was shown that
the source of the energy momentum tensor is the self-gravitating
non-linear magnetic monopole
\cite{Garcia1}.
We show that ultra-high energy collisions
can take place in this spacetime in a generic manner without
requiring any fine-tuning of the geodesic parameters and also
requiring only a finite proper time for such collisions to 
take place. 

We also investigate here the properties of the matter 
that might source such a geometry which has a regular center 
and hosts high 
energy collisions. We show that the conditions for the above 
and to have negative density gradient at the center imply 
the violation of strong energy condition near the center 
while the weak energy condition is still respected. Thus 
the center is surrounded by some sort of a dark energy fluid. 
Both the energy conditions can be shown to hold good at the point 
of collision. It would be interesting to investigate whether 
or not such a situation could arise in a realistic astrophysical 
scenario. Gravastars could possibly provide a scenario 
where such conditions could be met\cite{gravastar}.

In the next Section II, we consider the particle acceleration
in a general static spherical spacetime geometry which does not
contain any event horizon.
In Section III we discuss the energy conditions that must 
be satisfied by the matter for the regularity and for high 
energy collisions to occur. Section IV
considers an explicit example, which is the Bardeen
monopole geometry, in order to illustrate these considerations
and to see their applications. We finally conclude with a
discussion summarizing the implications of our results.

\section{Particle acceleration in general spherically symmetric
regular static spacetimes without an event horizon}
In this section we study the particle acceleration process
in the general spherically symmetric static spacetime whose metric is given by
\eq{E0}. A wide variety of spacetimes admit metric of this form. We derive the
conditions to have ultra-high energy particle collisions and
to have a regular center in the spacetime with no event horizon.
In other words, the spacetime has neither a black hole, nor
a naked singularity. 

We deal here with the spherically symmetric metric of the
general form,
\begin{equation}
 ds^2=-f(r)dt^2+f(r)^{-1}dr^2+r^2d\Omega^2
\label{E0}
\end{equation}
There is only one free function of the radial coordinate,
namely $f(r)$, in the metric and we shall impose the conditions
on it so as to have the above desired properties.

Since we are dealing with the spacetime that is static,
it admits a timelike Killing vector which is given by
$k=\partial_t$ and we have
\begin{equation}
 f(r)=-k.k
\label{E10}
\end{equation}
This implies that the free metric function $f(r)$ is
the negative of the norm of the timelike Killing vector $k$.
Thus the conditions that we impose on the metric function
$f(r)$ can be easily elevated to the coordinate independent
conditions to be imposed on the timelike Killing vector.

We further assume that the spacetime is asymptotically flat.
The norm of the timelike Killing vector is normalized to
$k.k=-f \rightarrow -1 $ at infinity as $r\rightarrow \infty$.

The condition for the existence of the event horizon
at any value of $r$ is given by $f(r)=0$.
Thus in order to avoid the horizon at any $r$ we must
have the following condition imposed on the metric function,
\begin{equation}
f(r)>0
\label{E8}
\end{equation}
namely, it must be nonzero and positive.

Now we derive conditions that must be imposed
on the metric function $f(r)$ so as to ensure the absence
of the spacetime singularity.
Essentially we must ensure that the curvature invariants
take a finite value everywhere.
We rewrite the function $f(r)$ as
\begin{equation}
 f(r)=1-\frac{2M(r)}{r}
\label{E9}
\end{equation}
where it has been traded for another function $M(r)$
of the radial coordinate, which is now the Misner-Sharp
mass for the system.
For the Schwarzschild spacetime this function is a
constant, $M(r)=M$, as a consequence of which there is
an event horizon at $r=2M$ and a singularity exists
at $r=0$. In our case we must therefore have
\begin{equation}
2M(r)<r
\label{E-1}
\end{equation}
in order to avoid the horizon. We must also have
$M(r)$ either tending to a constant value
or increasing slower than $r$ as we approach infinity
$r\rightarrow \infty$ so as to have $f(r)\rightarrow 1$.
We now derive
the conditions that must be imposed on $M(r)$ in
order to avoid the singularity.

Various curvature invariants can be written as
\begin{widetext}
\begin{eqnarray}
R=R_{\mu\nu}g^{\mu\nu}=\frac{4M'+2rM''}{r^2}\\
\nonumber
R_1=R_{\mu\nu}R^{\mu\nu}=\frac{8M'^2+2r^2M''^2}{r^4}
\end{eqnarray}
\begin{equation}
\nonumber
K=R_{\mu\nu\alpha\beta}R^{\mu\nu\alpha\beta}=\frac{4}{r^6}\left(12M^2+
4rM\left(-4M'+r M''\right)+r^2\left(8M'^2-4rM'M''+r^2M''^2\right)\right)
\end{equation}
\end{widetext}

All curvature invariants are finite at finite values
of the radial coordinate as long as the condition \eq{E-1}
for the absence of the horizon is satisfied.

However, various curvature invariants could diverge at the center $r=0$.
We Taylor expand $M(r)$ around $r=0$
\begin{equation}
 M(r)=M_{0}+M_{1}r+M_{2}r^2+M_{3}r^3+...
\end{equation}
For the curvature invariants to be finite at the center, it turns out we must have
\begin{equation}
 M_{0}=M_{1}=M_{2}=0
\label{E10}
\end{equation}
The first nonzero coefficient in the Taylor expansion must be $M_{3}$.
Thus the metric near center can be written as
\begin{equation}
 ds^2=-\left(1-2M_{3}r^2\right)dt^2+\left(1+2M_{3}r^2\right)dr^2+r^2d\Omega^2
\label{E5}
\end{equation}
If $M3\neq 0$ then the metric near the center looks like de sitter.

Equivalently the metric function $f(r)$ must satisfy the
following conditions,
\begin{equation}
 f(r=0)=1; f'(r=0)=0
\label{E7}
\end{equation}
Thus if the metric satisfies the conditions above
and if $f(r)$ takes a non-zero positive value, we can then
avoid singularity as well as event horizon in the spacetime.

We now derive the necessary and sufficient conditions
for the occurrence of ultrahigh energy collisions in spherically
symmetric static asymptotically flat spacetimes containing
no singularity or horizon.

Consider a particle which is following a geodesic
motion in the spherically symmetric static spacetime.
Let $U^{\mu}=\left(U^{t},U^{r},U^{\theta},U^{\phi}\right)$ be
the velocity of the particle. The motion of such a particle
would be confined to a plane which can be taken to be
the equatorial plane ($\theta=\frac{\pi}{2}$).
Thus
\begin{equation}
 U^{\theta}=0
\label{E1}
\end{equation}
Spherical symmetry and the static nature of spacetime
implies the existence of the Killing vectors
$k=\partial_t , l=\partial_{\phi}$. The following quantities
are then the constants of motion,
\begin{eqnarray}
 E=-\partial_t.U=f(r)U^{t}\\
\nonumber
L=\partial_{\phi}.U=r^2U^{\phi}
\end{eqnarray}
$E$ and $L$ can be interpreted as the conserved
energy and angular momentum per unit mass of the particle.
We can thus write the components of velocity as
\begin{eqnarray}
 U^{t}=\frac{E}{f(r)}
\label{E2}\\
\nonumber
U^{\phi}=\frac{L}{r^2}
\end{eqnarray}
Using normalization condition for velocity $U.U=-1$
and the \eq{E1},\eq{E2}, we can write the radial component
of the velocity $U^{r}$ as
\begin{equation}
 U^{r}=\pm \sqrt{E^2-f(r)\left(1+\frac{L^2}{r^2}\right)}
\label{E3}
\end{equation}
Here $\pm$ stands for the radially outgoing and ingoing
particles respectively.
This equation can be recast in the form
\begin{equation}
  U^{r2}+V_{eff}(r,L)=E^2
\end{equation}
where
\begin{equation}
 V_{eff}(r,L)=f(r)\left(1+\frac{L^2}{r^2}\right)
\end{equation}
can be thought of as an effective potential for
the motion in the radial direction.
The necessary condition for the particle to reach a
location with a radial coordinate $r$ is that the quantity
$E^2-V_{eff}(r,L)\ge 0$ must be nonnegative, which is
the condition for the radial velocity of the particle
to be real. If it is nonzero then the particle reaches
that point with a nonzero velocity. If it is zero then its
velocity is zero and it turns back. Whereas, if it is zero
and furthermore the potential also admits a maximum
$V_{eff}(r,L)'=0$ then the particle will asymptotically
reach that location at an infinite proper time.

We now consider a collision between two particles.
For the sake of simplicity we assume that the particles have
the same mass and that they also travel in the same plane
which is taken to be the equatorial plane. Let $m$ be the
mass and $U^1,U^2$ be the velocities of the particles.
The conserved energy and angular momenta are $E_1,E_2,L_1,L_2$.
The center of mass energy of collision between the particles
at the radial coordinate $r$ is given by
\begin{equation}
 \frac{E_{c.m.}^2}{2m^2}=1-U_{1}.U_{2}
\end{equation}
From \eq{E0},\eq{E1},\eq{E2},\eq{E3} we get
\begin{eqnarray}
  \frac{E_{c.m.}^2}{2m^2}=1+\frac{E_1 E_2}{f(r)}-k
\frac{|L_1||L_2|}{r^2}-\kappa \frac{1}{f(r)} \label{E4}\\
\nonumber
\sqrt{E_1^2-f(r)\left(1+\frac{L_1^2}{r^2}\right)}
\sqrt{E_2^2-f(r)\left(1+\frac{L_2^2}{r^2}\right)}
\end{eqnarray}
where $\kappa=1$ if either both the particles are radially
ingoing or they are outgoing.
However, if one of the particles is radially ingoing
and the other one is radially outgoing then we have
$\kappa=-1$. Whereas $k=1$ corresponds to the case where angular
momenta of both the particles have same sign and $k=-1$
corresponds to the case where angular momenta
of the particles have opposite signs.
We shall deal with these cases separately.

From \eq{E4} it is evident that there is a possibility
that the center of mass energy of collision can possibly
diverge when either $r=0$ or when $f(r)\rightarrow 0$ at some
finite value of the radial coordinate $r$.

We now show that the center of mass energy of collision
cannot diverge at the center of the spacetime $r=0$.
It is clear from \eq{E3} that a particle will reach the
center only if its angular momentum is zero.
Thus we must have $L_1=L_2=0$ in order for particles under
consideration to reach the center
and participate in the collision. And also it follows from
\eq{E2} that we must have $E_1,E_2 >0$. This is because absence of
the horizon implies $f(r)>0$ and the velocity
being future direction vector $U^t>0$.
The center of mass energy of collision between the particles
at the center using the fact $f(0)=1$ is then given by
\begin{equation}
  \frac{E_{c.m.}^2}{2m^2}=1+E_1E_2-\kappa \sqrt{E_1^2-1}\sqrt{E_2^2-1}
\end{equation}
which is clearly finite.

We now turn to the second case where $f(r)\rightarrow 0$
for some finite value of $r$.
It follows from \eq{E3} that for a particle with a given
energy $E$, the angular momentum must be in the following
range for it to reach $r$ and participate in the collision,
\begin{eqnarray}
\nonumber
L \in \left(-r\sqrt{\frac{E^2}{f(r)}-1},r\sqrt{\frac{E^2}{f(r)}-1}\right)
\label{E6}
\end{eqnarray}
Since $f(r)\rightarrow 0$ the limiting values of the allowed
angular momentum $L_{\pm}=\pm r\sqrt{\frac{E^2}{f(r)}-1}$ are extremely large.

We shall deal here with the following cases.
In the first case both the particles have finite radial velocity
at the point of collision.The
angular momenta of both the particles
are sufficiently away from the extreme critical values of
the allowed angular momentum interval.
In the second case both the particles take vanishingly
small value of radial component of velocity, but have nonzero
angular velocity at the collision.
The angular momenta of both the particles
are arbitrarily close to the limiting values.
In the third case the radial velocity of one the particles
if finite whereas the second particle has small radial component but finite
angular velocity at the point of collision.
The angular momentum of one of the particles is
sufficiently away from the limiting value and
that of the second particle is arbitrarily close to the limiting value.
In the fourth case one of the particles has both radial as well as angular
component of the velocity vanishing, whereas he second particle has either or both
components of the velocity non-vanishing at the collision.
In the first three cases both the particles have finite conserved energy.
Whereas in the fourth case the conserved energy of the first particle takes
a vanishing value, while second particle has finite conserved energy.

We note that of these, the first case is generic, whereas other 
cases require a fine-tuning.

{\it \bf Case 1}

We first focus on the case where the angular momenta of the
particles are well within the interval away from the limiting values and
conserved energies take finite values.
In that case the radial component of velocity at the point
of collision takes a finite nonzero value
and can be expanded in the following way, since the second
term under the square root
is much smaller than the first term,
\begin{equation}
 \sqrt{E^2-f(r)\left(1+\frac{L^2}{r^2}\right)}=
E\left(1-\frac{f(r)}{2E^2}\left(1+\frac{L^2}{r^2}\right)+...\right)
\end{equation}
The center of mass energy of collision \eq{E4},
neglecting the higher order terms can now be written as
\begin{eqnarray}
 \frac{E_{c.m.}^2}{2m^2}\approx 1+\left(1-\kappa\right)
\frac{E_1E_2}{f(r)}-k\frac{|L_1||L_2|}{r^2}\\
\nonumber
+\kappa\left(\frac{E_2}{2E_1}\left(1+\frac{L_1^2}{r^2}\right)\right)
+\kappa\left(\frac{E_1}{2E_2}\left(1+\frac{L_2^2}{r^2}\right)\right)
\end{eqnarray}

In the case where both the particles travel either
radially inwards or outwards {\it i.e.} when $\kappa=1$,
the center of mass energy of collision is given by
\begin{eqnarray}
\nonumber
\frac{E_{c.m.}^2}{2m^2} \approx 1-k\frac{|L_1||L_2|}{r^2}
+\frac{E_2}{2E_1}\left(1+\frac{L_1^2}{r^2}\right)
+\frac{E_1}{2E_2}\left(1+\frac{L_2^2}{r^2}\right)
\end{eqnarray}
which takes a finite value.

In the case where one of the particles under consideration
is radially ingoing and the other one is outgoing
{\it i.e.} when $\kappa=-1$ then the center of mass energy
of collision to the leading order can be written as
\begin{equation}
\frac{E_{c.m.}^2}{2m^2} \approx \frac{2E_1E_2}{f(r)}
\end{equation}
which clearly diverges in the limit $f(r)\rightarrow 0$.
Thus the center of mass energy of collision between an
ingoing and outgoing particles is arbitrarily large if the
metric function $f(r)$ computed at the point of collision
is arbitrarily close to zero.
\begin{equation}
 \lim_{f(r) \rightarrow 0} E_{c.m.}^{2} \rightarrow \infty
\end{equation}

Any particle satisfying \eq{E6} which is initially
radially ingoing will necessarily turn back as an outgoing particle
at an intermediate value of the radial coordinate between
the point of collision and the center if it has
a nonzero angular momentum. Whereas if the angular momentum is
zero, it will turn back as an outgoing particle
if $E<1$. If $E>1$ it will pass
through the center and then emerge as an outgoing particle.
Thus the origin of the outgoing particles which participate
in the collisions is easy to explain. The ingoing particles
either turn back or pass through the center and emerge as
the outgoing particles when their angular momentum for
any given value of energy lies in the range \eq{E6}.

{\it \bf Case 2}

We now turn to the second case where the angular momenta of both the particles
are arbitrarily close to the extreme values and the conserved energies 
take finite values. The radial velocity
of the particles is extremely small at the point of collision.
Angular momenta of the particles take a value which is given by
\begin{eqnarray}
 L_1 \approx \pm r\sqrt{\frac{E_1^2}{f(r)}-1}\\
\nonumber
 L_2 \approx \pm r\sqrt{\frac{E_2^2}{f(r)}-1}
\label{E30}
\end{eqnarray}
Since we are dealing with the case where $f(r)\rightarrow 0$, the angular
momenta take an extremely large value $L_1,L_2 \rightarrow \infty$.
Thus this situation is not generic and would occur only
under exceptional circumstances.
We assume that the angular momenta are fine-tuned to the
critical values to such an extent that
the contribution of the radial component of velocities to
the center of mass energy can be ignored.
\begin{equation}
 \sqrt{E_1^2-f(r)\left(1+\frac{L_1^2}{r^2}\right)}
\sqrt{E_2^2-f(r)\left(1+\frac{L_2^2}{r^2}\right)} <<f(r)
\label{E31}
\end{equation}
Thus it follows from \eq{E4},\eq{E30},\eq{E31} that
the center of mass energy of collision in this case is given by
\begin{equation}
 \frac{E_{c.m.}^2}{2m^2}\approx 1+\frac{E_1 E_2}{f(r)}\left(1-k\right)
+\frac{k}{2}\left(\frac{E_1}{E_2}+\frac{E_2}{E_1}\right)
\end{equation}
If the angular momenta of two particles are of the
same sign,
{\it i.e.} if $k=1$, then the center of mass energy of collision
turns out to be finite
\begin{equation}
 \frac{E_{c.m.}^2}{2m^2}\approx 1+\frac{1}{2}\left(\frac{E_1}{E_2}
+\frac{E_2}{E_1}\right)
\end{equation}
Whereas in the case where the orientation of the angular
momenta of the particles is opposite, {\t i.e.} when $k=-1$
the center of mass energy of collision is given by
\begin{equation}
 \frac{E_{c.m.}^2}{2m^2}\approx \frac{2E_1E_2}{f(r)}
\end{equation}
which  diverges in the limit $f(r)\rightarrow 0$.
Thus the center of mass energy of collision between the
particles which have nearly zero radial velocity,
but angular momenta with the opposite signs is arbitrarily
large in the limit where $f(r)\rightarrow 0$ at the point of collision.
\begin{equation}
 \lim_{f(r) \rightarrow 0} E_{c.m.}^{2} \rightarrow \infty
\end{equation}

{\it \bf Case 3}

We now discuss the third case where the angular momentum
of one of the particles is arbitrarily close to the extreme
value of the allowed
range of the angular momenta. The angular momentum of the
second particle is well within the allowed interval.
\begin{eqnarray}
 L_1 \approx \pm r\sqrt{\frac{E_1^2}{f(r)}-1}
\label{E33}
\end{eqnarray}
Both the particles have finite values of the conserved energy.
The radial velocity of the first particle takes a vanishingly
small value, whereas the radial velocity of
the second particle takes a finite value at the point of collision.
This situation is again not very generic.
The radial velocity of the first particle is assumed to be
so small that we can ignore its contribution to the center of mass
energy of collision.
\begin{equation}
 \sqrt{E_1^2-f(r)\left(1+\frac{L_1^2}{r^2}\right)}
\sqrt{E_2^2-f(r)\left(1+\frac{L_2^2}{r^2}\right)} <<f(r)
\label{E34}
\end{equation}
Thus from \eq{E4},\eq{E33},\eq{E34}, the center of mass energy
of collision to the leading order is given by,
\begin{equation}
 \frac{E_{c.m.}^2}{2m^2}\approx 1+\frac{E_1E_2}{f(r)}-k\frac{L_2E_1}{r\sqrt{f(r)}}
\end{equation}
In the limit where at the point of collision $f(r)\rightarrow 0$
we have,
\begin{equation}
 \frac{E_{c.m.}^2}{2m^2}\approx \frac{E_1E_2}{f(r)}
\end{equation}
which is clearly divergent.
Thus the center of mass energy of collision between
the particles, one of which has a vanishingly small radial velocity and
the other one has a finite radial component of velocity, diverges
in the limit where $f(r)\rightarrow 0$.
\begin{equation}
 \lim_{f(r) \rightarrow 0} E_{c.m.}^{2} \rightarrow \infty
\end{equation}
The center of mass energy of collision is s,smaller by the factor of $2$
as compared to the first two cases.

{\it \bf Case 4}

So far we have dealt with the situations where both the particles
have atleast one of the radial or angular component of velocity nonzero.
We now turn to the fourth case where one of the colliding particles
has vanishing radial as well as the angular components of velocity.
From \eq{E2},\eq{E3} it can be shown that
\begin{equation}
L_1 \approx 0 \; \ E_1 \approx \sqrt{f(r)} \approx 0
\label{E100}
\end{equation}
Thus both the angular momentum and the conserved energy of the particle
are approximately zero.
Second particle has a finite conserved energy. It has atleast one
of the radial or angular component nonzero.

The center of mass energy of collision between two such particles, using
\eq{E4},\eq{E100}, to the leading order is given by
\begin{equation}
 \frac{E_{c.m.}^2}{2m^2}\approx \frac{E_2}{\sqrt{f(r)}}
\end{equation}
which clearly diverges in the limit where
$f(r)\rightarrow 0$.
\begin{equation}
 \lim_{f(r) \rightarrow 0} E_{c.m.}^{2} \rightarrow \infty
\end{equation}
The divergence of the center of mass energy of collision when one of the
particles is at rest is however significantly slower
as compared to first three cases when both the particles
are in motion at the collision.

We have demonstrated in the above the divergence of the
center of mass energy of collision between the particles away
from the center at the radial
location where $f(r)\rightarrow 0$ in various cases.

The situation where both the particles have finite radial velocities
at the center is generic and would be of great interest from the perspective of
real physical scenarios. Whereas the other situations are not generic since
they require finetuning
of the angular momentum and energy of the particles as we discussed earlier.

We have shown that, in order to avoid the horizon we must have $f(r)>0$, whereas
to have ultrahigh energy collisions it is necessary that
$f(r)\rightarrow 0$. Also at infinity and at the regular center
we have $f(0)=f(r\rightarrow \infty)=1$.  Therefore it is necessary for the metric
function $f(r)$ to admit a minimum at an intermediate value of the
radial coordinate, with minimum value arbitrarily
close to zero, nevertheless positive,
\begin{equation}
 f'(r)=0; f(r)\rightarrow 0^{+}
\label{E9}
\end{equation}

For a static spherically symmetric asymptotically flat
spacetime considered here, the conditions imposed on $f(r)$,
namely \eq{E8},\eq{E7},and \eq{E9}, that are
required for avoidance of the singularity and horizon, and
in order to have ultrahigh energy collisions, can be translated
to the conditions to be imposed
on the timelike Killing vector $k=\partial_t$ making use of
\eq{E10}, and thus these can be written in a coordinate
independent way as follows:

I. In order to avoid the horizon, the timelike Killing vector
must retain its timelike nature
\begin{eqnarray}
k.k<0
\end{eqnarray}
This is also the condition to avoid singularity away from the center.

{II. In order to avoid the singularity we must have
\begin{eqnarray}
 \nabla_{\mu}(k.k)=0 ;
k.k=-1
\end{eqnarray}
at the center.

III. In order to have ultrahigh energy collisions
away from the center we must have
\begin{eqnarray}
 \nabla_{\mu}(k.k)=0;
k.k\rightarrow 0^{-}
\end{eqnarray}
That is, the norm of the timelike Killing vector must
admit a maximum, with the negative value at the maximum which is
arbitrarily close to zero.

Here we would like to note that since $f(r)$ admits a minimum
if the high energy collisions occur,
it implies that the gravity is repulsive in the region below
the radius at which the minimum occurs, upto the next maximum as we go
inside and $f(r)$ could in principle admit several minima.
If it admits only one minimum as
in the case of the example we shall discuss in the next section,
gravity is repulsive from the
center to the minimum and it is attractive from the minimum to
infinity. If $f(r)$ admits
several minima with extremely small minimum values then the high
energy collisions could occur
at several locations in the spacetime. Gravity is repulsive
between a maxima and minima and attractive between
a minima and maxima as we move from inside out.

It is also worthwhile to note that if not only the first
derivative but several derivatives of $f(r)$
take a zero value for some value of $r=r_a$,
{\it i.e.} $f^1(r_a)=f^{2}(r_a)=...=f^{n}(r_a)=0$ for large $n$,
where $f^{k}(r)$ stands for the $k^{th}$ derivative
with respect to $r$, then the value of $f(r)$ could be extremely
small in a finite interval around $r=r_a$,
where high energy collisions can take place.

\section{Energy conditions}
In this section we discuss the properties of the matter that 
would be required to source a regular spacetime geometry without 
an event horizon or naked singularity,
which can host ultrahigh energy collisions. Specifically, we 
investigate whether matter violates the energy conditions.

The energy momentum tensor for the metric \eq{E0} is given by
\begin{widetext}
\begin{equation}
 T_{\mu}^{\nu}=\frac{1}{8\pi}G_{\mu}^{\nu}=\frac{1}{8\pi}Diag
\left[\frac{-1+f+rf'}{r^2} \ , \
\frac{-1+f+rf'}{r^2} \ ,\ \frac{f'}{2}+\frac{f''}{2} \ ,\ 
\frac{f'}{2}+\frac{f''}{2}\right]\ =-\frac{1}{8\pi}
\ Diag\left[ \frac{2M'}{r^2} \ ,\ \frac{2M'}{r^2} \ , 
\ \frac{M''}{r} \ , \ \frac{M''}{r} \right]
\end{equation}
\end{widetext}

The density and pressures can be read from the expression above.
\begin{equation}
 \rho=-P_{r}=\frac{1}{8\pi}\frac{\left(1-f-rf'\right)}{r^2}=
\frac{1}{8\pi}\frac{2M'}{r^2}
\label{rhor}
\end{equation}
\begin{equation}
P_{\theta}=P_{\phi}=\frac{1}{8\pi}\left(\frac{f'}{2}+
\frac{f''}{2}\right)=-\frac{1}{8\pi}\frac{M''}{r}
\label{pt}
\end{equation}

The weak energy condition is satisfied if
\begin{equation}
\rho \ge 0  \label{we1}
\end{equation}
\begin{equation}
\label{we2}
 \left( \rho+P_{r} \right) \ge 0 \ , \  \left( \rho+P_{\theta} \right) \ge 0 \ , \
 \left( \rho+P_{\phi} \right) \ge 0
\end{equation}
whereas the strong condition is satisfied when
\begin{equation}
\left( \rho+P_{r}+P_{\theta}+P_{\phi} \right) \ge 0
\end{equation}
as well as \eq{we2} hold good.

From \eq{rhor},\eq{pt} we get
\begin{equation}
 \left( \rho+P_{r} \right) =0
\end{equation}
\begin{equation}
 \left( \rho+P_{\theta} \right)= \left( \rho+P_{\phi} \right)=
\frac{2M'-rM''}{r^2}=\frac{1-f}{r^2}+\frac{f''}{2}
\label{se1}
\end{equation}
\begin{equation}
\left( \rho+P_{r}+P_{\theta}+P_{\phi} \right)=-\frac{2M''}{r}=\frac{2f'}{r}+f''
\label{se2}
\end{equation}

As we have shown in the previous section the metric near 
the center is given by \eq{E5} and
the mass function to leading order is given by 
$M(r)=M_3r^3+M_4r^4+...$ .
It follows from \eq{we1} that for the energy density to be 
positive at the center we must have
\begin{equation}
M_3>0
\label{c1}
\end{equation}
From \eq{we1} we get
\begin{equation}
M_4=\pi \frac{d\rho}{dr}|_{r=0}
\end{equation}
If the energy density goes on decreasing as we go 
away from the center, which is reasonable to demand,
we must have
\begin{equation}
M_4<0
\label{c2}
\end{equation}
Near the center to the leading order we get
\begin{equation}
\left( \rho+P_{r}+P_{\theta}+P_{\phi} \right)=-12M_3 <0
\end{equation}
which imples that the strong energy condition is violated.
Whereas
\begin{equation}
  \left( \rho+P_{\theta} \right)= \left( \rho+P_{\phi} \right)=-4M_{4}r \ >0
\end{equation}
Thus the weak energy condition is satisfied.

Therefore if we assume that the density is positive and 
decreases away from the center, the weak
energy condition is satisfied but strong energy condition is 
violated in a region close to the
regular center. In this sense, the spacetime contains 
a ball of dark energy in the region around the center.

At the point of collision, as we showed in the previous section
$f=\epsilon\rightarrow 0^{+}$,$f'=0$ and $f''> 0$.
Thus it is clear from the \eq{se1},\eq{se2} that
\begin{equation}
 \left( \rho+P_{\theta} \right)=\frac{1-\epsilon}{r^2}+\frac{f''}{2}>0
\end{equation}
and
\begin{equation}
 \left( \rho+P_{r}+P_{\theta}+P_{\phi} \right)=f''\ge 0
\end{equation}
Thus both the weak energy condition and strong energy 
condition are satisfied at the point of collision.

We showed that the conditions for having a regular center 
and ultrahigh energy collisions in the spacetime imply that 
the regular center is surrounded by the dark energy fluid 
that violates the strong energy conditions and respects 
the weak energy conditions, whereas both the energy 
conditions are respected at the point of collision.

\section{An example: The Bardeen spacetime}
In the previous section we described, under what conditions,
ultrahigh energy collisions can take place in a class of
spherically symmetric static spacetimes containing neither
blackholes nor naked singularities. In this section we now
present a concrete example to illustrate this point.

We consider here the spacetime metric given by Bardeen
\cite{Bardeen}.
The metric can be written as
\begin{equation}
 ds^2=-\left(1-\frac{2mr^2}{(r^2+q^2)^{\frac{3}{2}}}\right)
dt^2+\left(1-\frac{2mr^2}{(r^2+q^2)^{\frac{3}{2}}}\right)^{-1}dr^2+
r^2d\Omega^2
\end{equation}
This metric takes the same form as \eq{E0} with
\begin{eqnarray}
 f(r)=\left(1-\frac{2mr^2}{(r^2+q^2)^{\frac{3}{2}}}\right)
\end{eqnarray}
and when compared with \eq{E9} we get
\begin{eqnarray}
M(r)=\frac{mr^3}{(r^2+q^2)^{\frac{3}{2}}}
\end{eqnarray}

There are two parameters in the solution, namely $m$ and $q$.
Here $m$ can be interpreted as the mass parameter for the
system. The other parameter $q$ can be interpreted
as the magnetic charge, as it was shown in
\cite{Garcia1}
that this metric solves Einstein equations
with a self-gravitating nonlinear magnetic monopole
as source. Various other examples of the regular blackholes 
have also been discovered and studied recently \cite{Garcia2}.

First of all $M(r)$ can be expanded in the
following way near the center $r=0$.
\begin{equation}
 M(r)=\frac{m}{q^3}r^3-\frac{3m}{2q^5}r^5+...
\end{equation}
This implies that $M_0=M_1=M_2=0$.
Thus it follows from \eq{E10} that the center is not
singular. Also, the spacetime is asymptotically flat.
We note that the function $f(r)$ admits only
one minimum at
\begin{equation}
r_{min}=\sqrt{2}q
\end{equation}
and the minimum value is given by
\begin{equation}
f(r_{min})=1-\frac{4}{3^{\frac{3}{2}}}\frac{m}{q}
\end{equation}

When $m>\frac{3^{\frac{3}{2}}}{4}q$ the minimum value
is less than zero, $f(r_{min})<0$.
Since $f(r=0)=f(r\rightarrow \infty)=1>0$, the function
$f(r)$ must admit a zero for two values of r,
$f(r)=0$ for $r=r_1,r=r_2$, where $0<r_1<r_{min}$ and
$r_{min}<r_2<\infty$.
The Bardeen metric in this case corresponds to the
blackhole with inner and outer horizons located at $r=r_1$
and $r=r_2$ respectively. The blackhole does not have a
singularity at the center and this was the first example
of a nonsingular blackhole.

When $m=\frac{3^{\frac{3}{2}}}{4}q$, the minimum value
of the function is zero, $f(r_{min})=0$, and both the zeros
coincide with the minimum
$r_{min}=r_{1}=r_{2}=0$. This corresponds to the case where
the inner and outer event horizons coincide. In such a case,
the Bardeen solution corresponds to an extremal
blackhole with a regular center.

When $m<\frac{3^{\frac{3}{2}}}{4}q$, the minimum of
the function is greater than the zero at $f(r_{min})=0$
and thus $f(r)$ is strictly positive everywhere in the
spacetime. There is then no event horizon or singularity
in the spacetime.

We are interested in the case where the mass parameter
is smaller than the value required for the minimum of
$f(r)$ to be zero,
\begin{equation}
 m=\frac{3^{\frac{3}{2}}}{4} q\left(1-\epsilon\right)
\end{equation}
where $0<\epsilon<1$.
The minimum value of the function $f(r)$ at $r=q$
is given by
\begin{equation}
f(r_{min})=\epsilon
\label{E11}
\end{equation}
As discussed before, at the center and at infinity
this function is unity and it admits only one extremum,
namely minimum at $r=q$. Thus it is everywhere positive.
The solution is regular and admits no horizons.

We now describe high energy particle collisions.
For the sake of simplicity and definitiveness we restrict
our attention to
two particles that move along radial direction along the same line.
The angular momenta of these particles are zero $L_1=L_2=0$.
We further assume that $E_1=E_2=E$.
The radial component of velocity from \eq{E3} in
this case is given by
\begin{equation}
 U^r=\pm \sqrt{E^2-f(r)}
\end{equation}
It follows that we must have $E\ge \sqrt{\epsilon}$.
If $E=\sqrt{\epsilon}$ then particle stays at rest at $r=r_{min}=\sqrt{2}q$
If $\sqrt{\epsilon}\le E<1$ then an initially ingoing particle will turn back
at a radial coordinate between the center and the minimum
and emerge as an outgoing particle.
This implies that the gravity is "repulsive" close
to the center. This is quite surprising that gravity
can be repulsive in the absence of the naked singularities.
If the particle has energy $E>1$ then initially ingoing
particle passes through the center and emerges as an outgoing
particle from the
other side, whereas if $E=1$ the ingoing particle
asymptotically approaches the center.

We can now take two particles each with $E>\sqrt{\epsilon}$ and $E\neq 1$.
An initially ingoing particle will emerge as an outgoing particle
either after getting reflected back or after passing through the center.
The center of mass energy of collision between the ingoing and
outgoing particles, using \eq{E4} is given by
\begin{equation}
\frac{E_{c.m.}^2}{2m^2}=\frac{2E^2}{f(r)}
\end{equation}
The center of mass energy is maximum where $f(r)$ is minimum. From \eq{E4}
we get,
\begin{equation}
\frac{E_{c.m.}^2}{2m^2}=\frac{2E^2}{\epsilon}
\end{equation}
The center of mass energy of collision can be arbitrarily
large in the "near-extremal" limit where $\epsilon \rightarrow 0$
\begin{equation}
\lim_{\epsilon \rightarrow 0}\frac{E_{c.m.}^2}{2m^2}
=\frac{2E^2}{\epsilon}\rightarrow \infty
\end{equation}

We now consider a collision between a particle which is at rest
at $r=r_{min}$, {\it  i.e} the particle with energy
$E_1=\sqrt{\epsilon}$ and zero angular momentum, with another radially
ingoing particle with energy $E_2$ which has a finite radial
velocity. It follows from \eq{E4} that the center of mass energy of
collision between the particles is given by
\begin{equation}
\frac{E_{c.m.}^2}{2m^2}=1+\frac{E_2}{\sqrt{\epsilon}}
\end{equation}
This clearly diverges in the limit $\epsilon \rightarrow 0$,
\begin{equation}
\lim_{\epsilon \rightarrow 0}\frac{E_{c.m.}^2}{2m^2}
\approx \frac{E_2}{\sqrt{\epsilon}}\rightarrow \infty
\end{equation}

We plot the quantities $\left( \rho+P_{\theta} \right)$ and $\left( \rho+P_{r}+P_{\theta}+P_{\phi} \right)$
as a function of radius in Fig1. It can be clearly seen that the weak energy condition is respected 
everywhere, whereas the strong energy condition is violated in the region near center upto a certain radius 
below the place where collision takes place. 
The violation in the regular blackholes was discussed in \cite{Zawa}

\begin{figure}
\begin{center}
\includegraphics[width=0.5\textwidth]{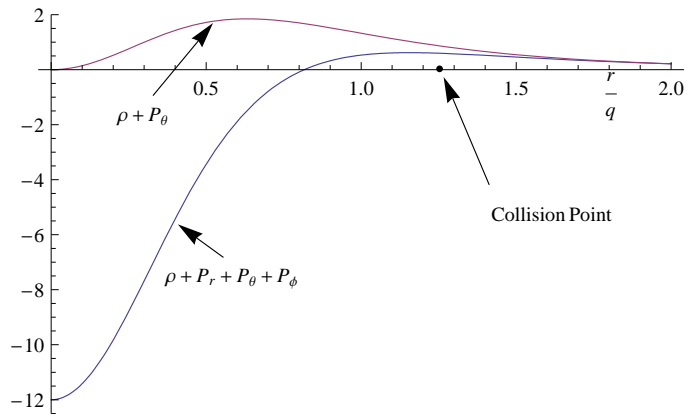}
\caption{\label{fg1}
In this figure we plot $\left( \rho+P_{\theta} \right)$ and
$\left( \rho+P_{r}+P_{\theta}+P_{\phi} \right)$
as a function of $x=\frac{r}{q}$ in
units of $\frac{m}{4\pi q^3}$. The weak energy condition
is satisfied everywhere.
Whereas the strong energy condition is violated in a ball around
the center with the radius smaller than the location of the collision.
It is respected everywhere else.
}
\end{center}
\end{figure}

Thus we have demonstrated here the ultrahigh energy
collisions of the particles in an explicit example of a spacetime
geometry which contains no event horizons or a naked singularity.

\section{Conclusions}
We showed here that it would be possible to have ultrahigh
energy particle collisions
in a spacetime which does not contain any event horizon or a naked
singularity. Previously, the phenomenon of ultrahigh energy collisions
had been explored in the blackhole geometries. We extended the
result to geometries containing naked singularities, and further
we find here that it is not necessary to have horizon or
singularities for high energy collisions to occur.

We obtained the conditions to be imposed on the general
spherically symmetric static spacetime in order to avoid the horizon
and singularities and in order to have ultrahigh energy collisions.
We showed that it order to avoid the horizon the norm of the timelike
Killing vector $k=\partial_t$ should remain negative everywhere.
This condition also ensures that there is no singularity away from the center.
It order to avoid the singularity at the center, the norm of the
Killing vector should admit a minimum and must take a value $-1$.
It order to have ultrahigh energy collisions between the particles
it is necessary for the norm of the timelike Killing vector to admit
a maximum away from the center with the value very close to zero.
If these conditions are met then the center of mass energy between
the radially ingoing and outgoing particles (with finite radial
component of velocity, which is the situation that would occur
generically), is arbitrarily large depending on how close to zero
is the norm of the timelike Killing vector. This is the condition
for the existence of the high blueshift surface for the approaching
particles and high redshift surface for the receding particles.

The conditions for having a regular center and ultrahigh energy collisions
in the spacetime imply that the strong energy condition is 
violated in the region surrounding the center, 
whereas the weak energy condition is respected. Thus the region 
near the regular center is filled with dark energy fluid. Both the 
energy conditions are respected at the location where high energy 
collisions take place.

We demonstrate this point here with a concrete example of the Bardeen
spacetime which corresponds to a perfectly regular geometry without
any horizon or singularities for a certain parameter range. We show
that the conditions mentioned above are met with and it is possible to
have ultrahigh energy collisions of particles in the Bardeen spacetime
for appropriate values of parameters.

The purpose of this paper was to show that unlike the common belief,
particle acceleration and ultrahigh energy collisions can take place
in a perfectly regular spacetime without blackholes or naked 
singularities. Some of the energy conditions must be, however, 
violated by the matter fields in such a case near the center. In 
other words, in a fully energy conditions preserving spacetime, 
either a blackhole or a naked singularity becomes a necessary 
condition for the high energy collisions to take place.   
In a future work we would like to generalize these results to
more general spacetime geometries.

\section{Acknowledgements}
We would like to thank K. Nakao and M. Kimura for discussions and
going though the manuscript carefully during their recent visit to TIFR.
MP would like to thank T. Harada for interesting discussion during
the ICGC meeting in Goa.


\begin{thebibliography}{99}


\bibitem{BSW}M.Banados, J.Silk, S.M.West, Phys. Rev. Lett. 103, 111102 (2009).

\bibitem{Berti}E.Berti, V.Cardoso, L.Gualtieri, F.Pretorius, U.Sperhake,
Phys. Rev. Lett. 103, 239001 (2009);
{Jacobson} T. Jacobson, T.P.Sotiriou, Phys. Rev. Lett. 104,
021101 (2010).

\bibitem{Harada}T. Harada, M.Kimura, Phys. Rev .D 84, 124032(2011);
Phys .Rev . D 83, 024002(2011);Phys .Rev . D 83,084041(2011);
A. Grib, Y.Pavlov, Grav. Cosmol.17, 42,(2011),arXiv:1007.3222 (2010);
arXiv:1004.0913(2010),M. Bejger, T. Piran, M. Abramowicz, F. Håkanson
arXiv:1205.4350 (2012).
\bibitem{BSW2}M.Banados, B.Hassanain, J.Silk, S.West,
Phys. Rev. D 83, 023004,(2011);A. Williams, Phys. Rev .D 83, 123004,(2011).
\bibitem{Nakao}M.Kimura, K.Nakao, H.Tagoshi, Phys. Rev . D 83, 044013,(2011).
\bibitem{Others}T. Igata, T.Harada, M. Kimura, arXiv:1202.4859 (2012);
J. Yang, Y. Li, Y. Li, S. Wei, Y.Liu, arXiv:1202.4159 (2012)[hep-th]; [hep-th]
V. Frolov, arXiv:1110.6274 (2011);
O. B. Zaslavskii, Phys. Rev. D 85 (2012) 024029; arXiv:1105.0303
(2011);
Phys .Rev . D 84, 024007(2011);
JETP Lett. 92, 571 (2010); arXiv:1107.3964 (2011);
Class .Quant .Grav .28 ,105010, (2011); Phys. Rev. D82, 083004,(2010).
C. Zhong, S. Gao, JETP Lett. 94, 8,589 (2011);
J. Sadeghi, B. Pourhassan, arXiv:1108.4530 (2011);
Y. Zhu, S. Wu, Y. Jiang, G. Hong Yang, Phys. Rev. D 84, 123002 (2011);
A. Abdujabbarov, B. Ahmedov, B. Ahmedov, Phys. Rev. D, 84, 4, 044044 (2011);
S. Gao, C. Zhong , Phys. Rev. D84, 044006,(2011);
W. Yao, S. Chen, C. Liu, J. Jing, Eur. Phys. J. C 72, 1898 (2012);
J.Said,  K. Adami, Phys. Rev. D83 ,104047,(2011);
A. Grib, Y. Pavlov, O. Piattella, arXiv:1105.1540 (2011);
C. Liu, S. Chen, J. Jing, arXiv:1104.3225 (2011);
Y. Zhu, S. Wu, Y. Xiao Liu, Y. Jiang, Phys. Rev. D84, 043006,(2011);
C. Liu, S. Chen, C. Ding, J. Jing, Phys. Lett. B 701, 285, (2011);
Y. Li, J. Yang, Y. Li, S. Wei, Y. Liu, Class. Quantum Grav. 28 225006 (2011);
R. Plyatsko, O. Stefanyshyn, M. Fenyk, Phys. Rev. D 82, 044015 (2010);
P. Mao, R. Li, L. Jia, J. Ren, arXiv:1008.2660 (2010);
S. W. Wei, Y. X. Liu, H. T. Li, and F. W. Chen, JHEP 1012, 066 (2010);
S. W. Wei, Y. X. Liu, H. Guo, and C. E Fu, Phys. Rev.D 82, 103005 (2010).




\bibitem{Patil}M. Patil, P. Joshi, Class. Quantum Grav. 28,
235012 (2011); Phy. Rev. D 84, 104001 (2011).
\bibitem{Patil2}M. Patil, P. Joshi, K. Nakao, M. Kimura arXiv:1108.0288(2011).
\bibitem{Patil3}M. Patil, P. Joshi,arXiv:1106.5402(2011).
\bibitem{Patil4}M. Patil, P. Joshi,arXiv:1112.2525(2011).
\bibitem{Bardeen}J. Bardeen, presented at GR5, Tiflis, U.S.S.R., and published
in the conference proceedings in the U.S.S.R., 1968
\bibitem{Garcia1} E.Ayon-Beato, A. Garcia, Phys. Lett. B 493 149 (2000)
\bibitem{gravastar}P. Mazur, E. Mottola  arXiv:gr-qc/0109035 (2001);
M. Visser, D.Wiltshire, Class.Quant.Grav. 21 (2004) 1135-1152;
C. Ghezzi, Astrophys.Space Sci.333:437-447,(2011);
R. Chan, M. Silva, P. Rocha, A. Wang, JCAP 0903:010,(2009);
JCAP 0811:010,2008;
J. Lemos, O. Zaslavskii,Phys.Rev.D78:024040,(2008).
\bibitem{Garcia2}E.Ayon-Beato, A. García, Phys.Rev.Lett.80:5056-5059,(1998);
Gen.Rel.Grav. 31 (1999) 629-633;Phys.Lett. B464 (1999) 25.

\bibitem{Zawa} O. B. Zaslavskii, Physics Letters B 688 (2010) 278-280
\end{thebibliography}
\end{document}